\documentclass[a4paper,11pt]{article}
\usepackage{jinstpub} 
\usepackage{comment}
\usepackage[rawfloats=true]{floatrow}
\usepackage{siunitx} 
\DeclareSIUnit\sq{\ensuremath{\Box}}                           

\title{\boldmath Development of non amplified Depleted MAPS sensors towards 50 ps timing resolution on charged particles}

\author[a,b]{P.~Schwemling} \note{Corresponding author.}
\author[c,d]{, R. Casanova}
\author[a]{, Y.~Degerli}
\author[c,e]{, Y. Gan}
\author[c,f]{, S. Grinstein}
\author[a]{, F.~Guilloux}
\author[g]{, T. Hemperek}
\author[e]{, G. Huang}
\author[a]{, JP.~Meyer}

\affiliation[a]{IRFU, CEA, Universit\'{e} Paris-Saclay, F-91191 Gif-sur-Yvette, France}
\affiliation[b]{ Université Paris-Cité, Campus Grands Moulins, 75013, Paris, France}
\affiliation[c]{Institute for High Energy Physiscs (IFAE)
 Campus Autonomous University of Barcelona (UAB), 08193, Bellaterra (Barcelona), Spain}
 \affiliation[d]{Department of Microelectronics and Electronic Systems, Autonomous University of Barcelona (UAB),
 Campus Autonomous University of Barcelona (UAB), 08193, Bellaterra (Barcelona), Spain}
 \affiliation[e]{PLAC, Key Laboratory of Quark and Lepton Physics (MOE), Central China Normal University,
 152 Luoyu Road, Wuhan, 430079, Hubei, China}
 \affiliation[f]{Catalan Institution for Research and Advanced Studies (ICREA)}
 \affiliation[g]{DECTRIS AG, Taefernweg 1, 5405 Baden-Daettwil, Switzerland}

\emailAdd{philippe.schwemling@cea.fr}

\abstract{The MiniCactus sensors are demonstrator sensors designed in LFoundry LF15A \SI{150}{nm} technology, intended to study the performance of non amplified High Voltage High Resistivity CMOS sensors for measurement of time of arrival of charged particles. This paper presents the context, design features and some of the first test-beam results obtained with the latest MiniCactus sensor version, MiniCactus V2. With a
\SI{175}{\micro m} thick sensor biased at -\SI{350}{V}, we have obtained a \SI{60}{\pico s} time resolution on Minimum Ionizing Particles detected with a \SI{500}{\micro m} $\times$ \SI{500}{\micro m} pixel.}


\begin{document}
\maketitle
\flushbottom

\section{Introduction}
Measurement of the time of arrival of Minimum Ionizing Particles (MIPs) is now recognized as an important tool to
disentangle individual collisions in high pile-up environments, such as those found at the high-luminosity LHC or at the future FCC-hh
collider.

On the longer term, other technologies than the presently largely used LGADs have to be investigated, to drive down the cost of sub detector systems 
potentially featuring hundreds of square meters of sensor surface.
Measurement of the time of arrival of charged particles with a resolution of the
order of \SI{20}{\pico s} is also one of the strategic goals of the CERN DRD3 collaboration.

One very promising technology to reach this goal is High Voltage High Resistivity CMOS ~\cite{hv-cmos}. This kind of sensor has the advantage
of being relatively cheap, since it is based on high production volume industrial processes. It is technically
possible to integrate on the same substrate the sensor diode, the analog front-end processing electronics, shaper and
discriminator, and at least part of the digital processing electronics, amplitude, time of arrival measurement and data
serialisation. With such a sensor, there is no need for costly bump bonding operations.

\section{Sensor design}

\subsection{Design compromises}
Designing a non amplified monolithic timing sensor involves finding compromises on several parameters.
Large collecting electrode dimensions are favorable for field uniformity, but give higher noise. Thin sensors allow to
reduce the influence of the energy deposit fluctuations along the particle path, but reduce the signal over noise ratio, which increases
the jitter contribution (equal to the signal over noise divided by the rise time) to the total time resolution. The analog front-end electronics needs to have an 
impulse response with a rise time comparable or lower than the rise time of the physics current signal, which is a few ns. Achieving this typically means high bias currents and hence high power dissipation in the analog front-end. Power dissipation of several Watts per square centimeter is not a problem for small scale
demonstrators, but practical sensor implementation in actual experiments has to be compatible
with the available cooling power, which is at most a few hundreds of \SI{}{\milli W} per square centimeter. 

\subsection{MiniCactus V2 sensor design}   
MiniCactus V2 is a timing sensor intended to explore the compromises outlined in the previous section, and
to investigate the timing performance that can be reached with a large-electrode
sensor design implemented with a high-voltage high resistivity process, without intrinsic amplification of the collected
charge, in contrast to LGADs. MiniCactus V2 is also an improvement over the previous MiniCactus V1 chip, that had the
same goal. MiniCactus V1 test-beam results have been published in~\cite{minicactus_v1}. The best timing performance reached by MiniCactus V1 is a
65 ps time resolution on MIPs (180 GeV/c muons at the CERN H4 beamline), with a \SI{200}{\micro m} thick chip, on a
\SI{500}{\micro m} $\times$ \SI{1000}{\micro m} pixel operated at a bias voltage of 450 V~\cite{minicactus_v1}.

The technology used for MiniCactus V1 and MiniCactus V2 is the LFoundry LF15A 150 nm process. This analog and digital mixed signal
technology provides up to six metal
layers, and the foundry gives the possibility to use High Voltage High Resistivity (2 k $\Omega$cm) wafers. After fabrication, the wafers
have to be thinned to the desired thickness. The backside of the wafers is then doped with Boron and metallized,
before finally being diced. Chips from wafers thinned down to 150, 175, and \SI{200}{\micro m} are available. MiniCactus V1 results suggest
the optimal thickness should lie in this range.

\begin{figure}[h]
\centering
  \floatsetup{heightadjust=all, valign=c}
  \begin{floatrow}
  \ffigbox{%
    \includegraphics[width=0.4\textwidth, keepaspectratio]{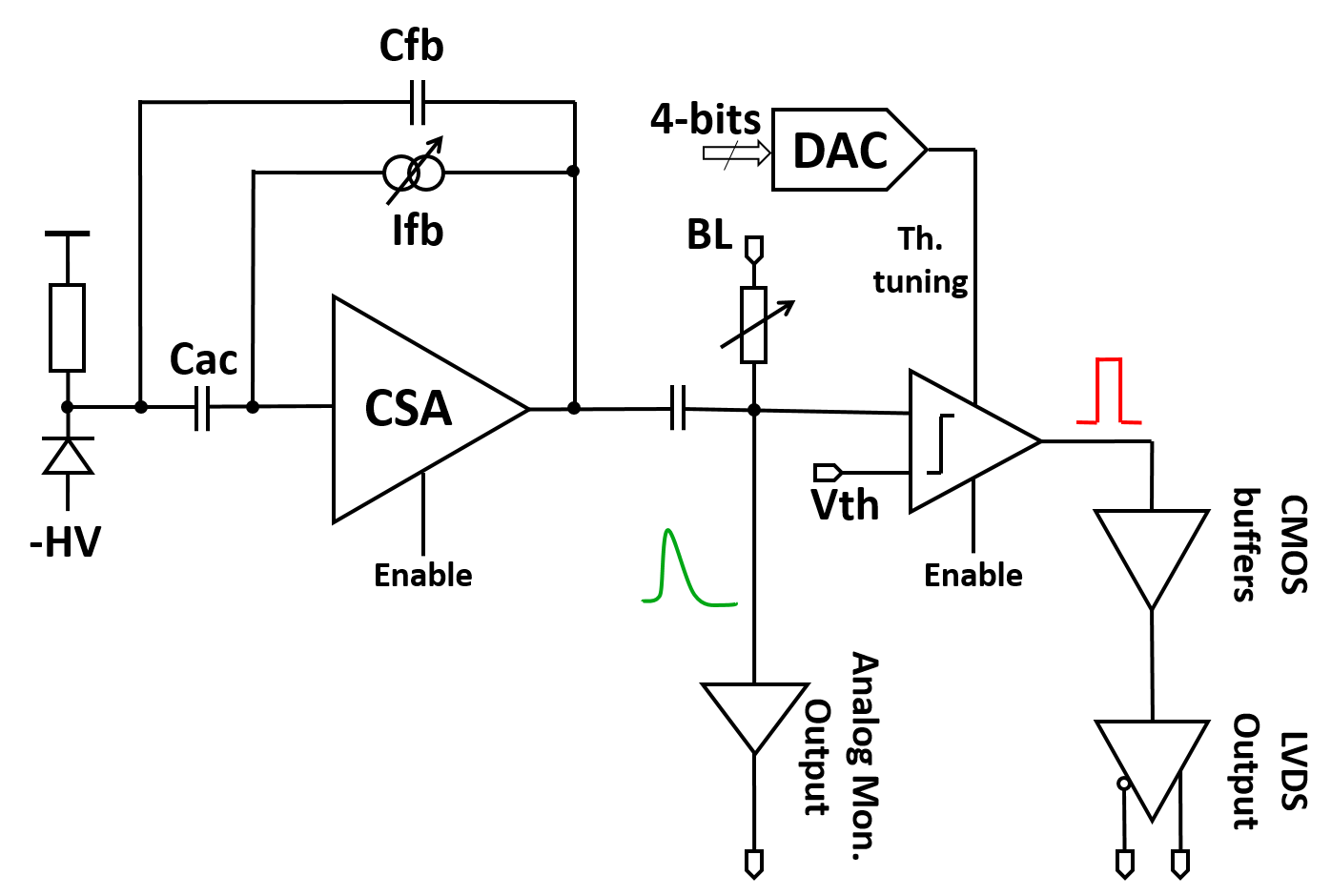}} {
    \caption{Front-end architecture. The analog preamplifier is on the left, the discriminator on the right.}
    \label{FE}}
  \ffigbox{%
    \includegraphics[width=0.4\textwidth,  keepaspectratio]{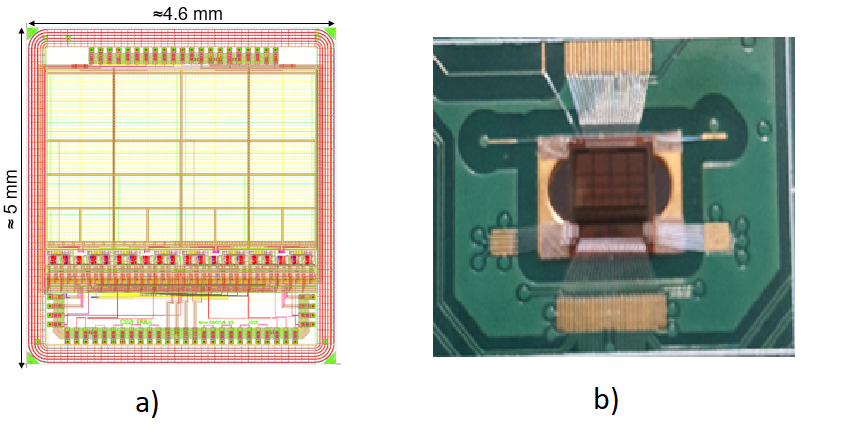}} {
    \caption{Left : Layout of the MiniCactus v2 chip. Right : Photograph of a MiniCactus v2 chip bonded to a PCB.}
    \label{layout-MiniCactus-v2}}
  \end{floatrow}
 \end{figure}

The typical architecture of the front-end and discriminator electronics attached to each sensor diode is shown in figure~\ref{FE}.
MiniCactus V1 uses only a charge sensitive preamplifier as analog front-end. This block, called CSA1, also used by MiniCactus V2, is an evolution of the preamplifier used
by the LF-CPIX chip~\cite{LF-CPIX1,LF-CPIX2}, after optimization for time measurement.

The MiniCactus V2 global layout is shown in figure~\ref{layout-MiniCactus-v2}. The chip is approximately 5 mm high $\times$ 4.6 mm wide. It is organized in four independant columns. Within each column, the MiniCactus V2
output pads are connected to small groups of two or four pixels. The slow control logic allows to activate any individual pixel within each group. Activated pixels send a discriminated LVDS digital signal, generated by the leading edge discriminator implemented in the front-end circuitry for each pixel (see figure~\ref{FE}), and an analog monitoring signal that is used to measure the signal amplitude. The discriminator is connected directly to the analog front-end output, the analog monitoring signal is a buffered copy of the discriminator input. Due to the limited bandwidth of the buffer, the monitoring signal is a somewhat slowed down copy of the discriminator input.  
In total, it is possible to have up to 8 pixels simultaneously active on the entire MiniCactus V2 chip. 
Each column has pixels of \SI{1000}{\micro m} $\times$ \SI{1000}{\micro m}, 500 $\times$ \SI{1000}{\micro m},
and \SI{500}{\micro m} $\times$ \SI{500}{\micro m}. There are also a few small diodes with dimensions \SI{50}{\micro m} $\times$ \SI{150}{\micro m} and \SI{50}{\micro m} $\times$ \SI{50}{\micro m}. The front-end parameters
have been optimized for capacitances corresponding to the larger diodes of dimensions \SI{500}{\micro m} $\times$ \SI{500}{\micro m} and above, the small diodes having to
be considered as test structures. The layout of all the sensor diodes of MiniCactus V2 is the same as the layout identified with the
MiniCactus V1 results as giving the best performance.

The front-end circuitry and slow control logic are implemented in the end of the columns
(figure~\ref{layout-MiniCactus-v2}). The slow control logic allows to configure discriminator thresholds, activate or veto individual pixels, 
set per pixel bias and feedback currents through four or five bits DACs, the DACs being part of the on-chip logic.

The main differences between MiniCactus V2 and MiniCactus V1 are the following :
\begin{itemize}
\item Since significant coupling from the digital has been observed on MiniCactus V1 \cite{minicactus_v1}, separation has been improved between the analog and digital signal paths on MiniCactus V2, by implementing the output LVDS drivers in a separate deep N-well, by shortening the paths of the CMOS digital signals, and adding a programmable hysteresis to the discriminator,
to avoid as much as possible spurious retriggering induced by noise or residual couplings.
\item Two new optimized analog front-ends have been implemented \cite{minicactus_v2}. The first one, called CSA2, is a charge sensitive amplifier with reoptimized
parameters. The second, called VPA, is a voltage sensitive preamplifier.
These two front-ends provide according
to Cadence simulations timing performance slightly better by a few picoseconds than the original CSA1. More important, they have a return
time to baseline less than \SI{25}{\nano s}, much smaller than the \SI{80}{\nano s} or so of CSA1. This makes them a possible drop-in replacement of LGADs for post
High Luminosity phase upgrades of LHC subdetectors, like the ATLAS HGTD.
\end{itemize}

The two leftmost columns of MiniCactus V2 are equipped with the CSA1 front-end, the third one is equipped with the CSA2 front-end, and the fourth column is equipped with the VPA front-end.

\section{Test setup}
The chips are bonded to a dedicated \SI{100}{\milli m} $\times$  \SI{100}{\milli m} PCB carrying all the ancillary electronics needed to bring high voltage, configure the chip, and
buffer all the analog and digital signals before routing them to the readout and data acquisition system. A photograph of a bonded
MiniCactus V2 chip is shown in figure~\ref{layout-MiniCactus-v2}.

First tests of MiniCactus V2 have taken place during summer 2024 at CERN, at the H4/EHN1 beamline, in parasitic mode together with the DRD1 collaboration.
The measurements were done using 180 GeV/c muons. The data has been acquired using a LeCroy HDO4104A 12 bits, 4 channels, \SI{1}{\giga Hz} bandwidth oscilloscope. The test setup has been described in detail
in~\cite{minicactus_v1}. Events have been acquired in triple coincidence between the MiniCactus v2 pixels, a Hamamatsu H11934 PMT coupled to a \SI{5}{\milli m} thick NE110
scintillator slab, yielding a resolution on MIPs of the order of \SI{60}{\pico s}, and the DRD1 beamline MCP coupled to a Cerenkov radiator. This configuration allows an independent determination of the time resolution of the three individual
devices, run by run.
\begin{figure}[h]
\centering
  \floatsetup{heightadjust=all, valign=c}
  \begin{floatrow}
  \ffigbox{%
    \includegraphics[width=0.55\textwidth, keepaspectratio]{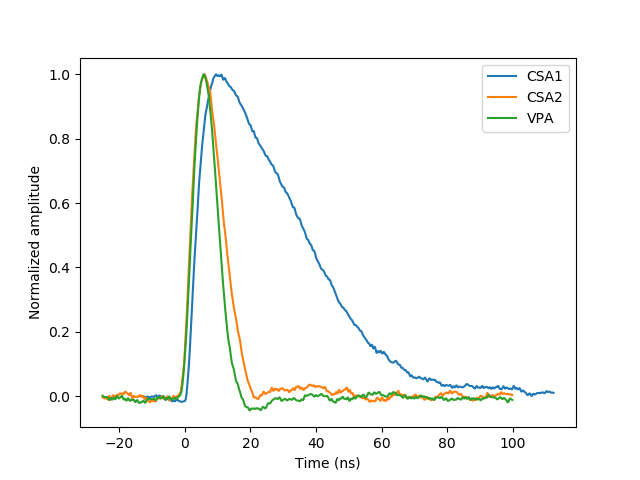}} {
    \caption{Normalised pulseshapes given by the three front-end architectures. The CSA1 has a return time to baseline of about 80 ns, CSA2 and VPA both return to baseline in about 20 ns.}
    \label{pulseshapes}}
  \ffigbox{%
    \includegraphics[width=0.8\textwidth,  keepaspectratio]{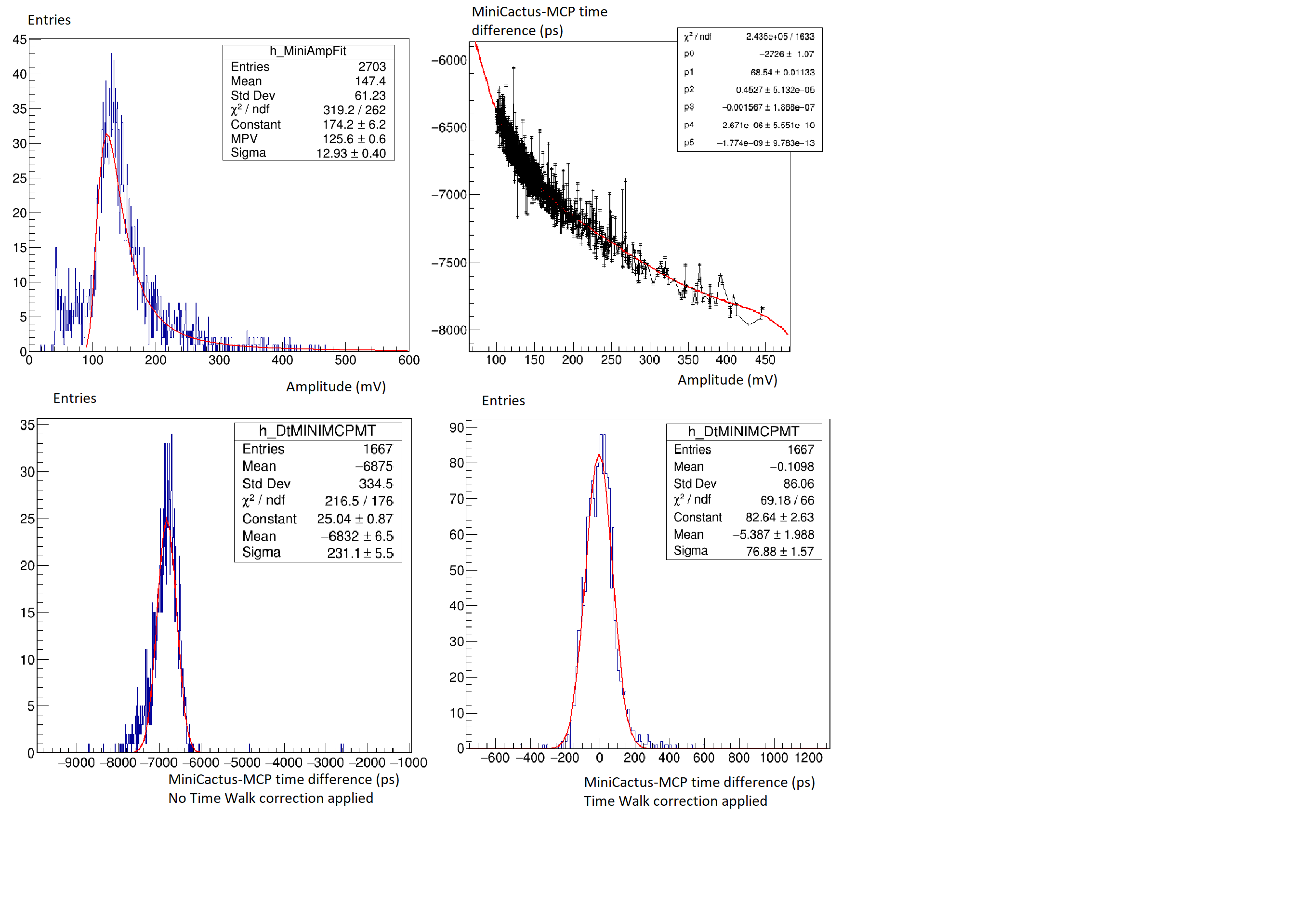}} {
    \caption{Data obtained from a \SI{500}{\micro m} $\times$ \SI{500}{\micro m} pixel polarized at \SI{-350}{V}, for a \SI{150}{\micro m} thick device. Upper left : Amplitude distribution. Upper right : correlation between amplitude and time measurement (Time walk correction). Lower left : MiniCactus and MCP time difference, before time walk correction. Lower right : MiniCactus and MCP time difference, after application of the time walk correction.}
    \label{tbeam-plots}}
  \end{floatrow}
 \end{figure}
\section{Test beam results}
Typical pulseshapes are shown in figure~\ref{pulseshapes}. The difference in return time to the baseline is clearly visible among the different front-end flavours,
and is as expected below \SI{25}{\nano s} for the pixels equipped with the CSA2 and VPA front-ends.
The amplitude spectrum and time resolution has been investigated for the different pixel sizes and front-end flavours. An example amplitude spectrum is shown in figure~\ref{tbeam-plots} (upper left). This spectrum follows as expected a Landau distribution.
For each pixel, events have been selected online requiring the analog monitoring signal amplitude to be \SI{20}{\milli V} above
the baseline. Further selection has been applied offline, requiring the presence of a signal in the PMT channel as well
as in the MCP channel. The time of arrival of the MCP and PMT signals has been measured event by event from their
digitized analog pulseshape, by computing the time at which the pulseshape reaches 10\% of its maximum above the pedestal.
For the MiniCactus signal, the time of arrival has been estimated from the discriminator output, by computing the time
at which the digital signal reaches 50\% of its maximum above the pedestal. Then, the distribution of the average time
difference between the MCP and the MiniCactus as a function of the MiniCactus analog monitoring amplitude has been fitted
to a fifth order polynomial, yielding the time walk correction function (figure~\ref{tbeam-plots} (upper right). The time walk correction function has been applied
event by event to the time of arrival of the MiniCactus digital signal, which allows to compute the final time resolution. Figure~\ref{tbeam-plots} (lower left) shows the distribution of the difference of time of arrival between the MiniCactus and the MCP, before application of the time walk correction. Figure~\ref{tbeam-plots} (lower right) shows the distribution of the difference of time of arrival between the MiniCactus and the MCP, after application of the time walk correction. The individual time resolution of the three devices in coincidence (MiniCactus, MCP, PMT) is extracted from the r.m.s. of the time difference distributions given the three device pairs. Under the same high voltage bias conditions and with the same pixel dimensions, the three different front-ends yield very close time resolutions. The best time resolution (59.5 $\pm$ 1.8 ps) has been obtained on a \SI{500}{\micro m} $\times$ \SI{500}{\micro m} pixel biased at -350 V, equipped with the CSA1 frontend. The power dissipation is around \SI{1.6}{\milli W} per front-end channel, i.e. 0.6 W.cm$^{-2}$ for a matrix equipped with \SI{500}{\micro m} $\times$ \SI{500}{\micro m} pixels.

\section{Conclusion}
MiniCactus V2, a sensor designed in LF15A 150 nm technology, and optimised for the measurement of the time of arrival of charge particles in the context of high energy physics experiments, has undergone a first testbeam campaign at CERN over summer 2024. The best time resolution, 60 ps, has been obtained on a \SI{500}{\micro m} $\times$ \SI{500}{\micro m} pixel for a \SI{175}{\micro m} thick chip biased at -350 Volts.

\acknowledgments
This project has received funding from the European Union's Innovation programme under grant no 101004761 (AIDAInnova). It has also received funding from the P2I department of Université Paris Saclay, and was also partially supported by MICIIN (Spain) with funding from European Union NextGenerationEU(PRTR-C17.I1), and from the Generalitat de Catalunya.

The authors would finally like to acknowledge the DRD1 team for their kindful help during the testbeam campaign at SPS-CERN.

\end{document}